\documentstyle[aps,pre,floats,epsfig]{revtex}

\begin{document}

\twocolumn[\hsize\textwidth\columnwidth\hsize\csname@twocolumnfalse%
\endcsname
\title{Persistence exponents in a 3D symmetric binary fluid mixture}

\author{V. M. Kendon$^{1,*}$, M. E. Cates$^1$, J-C. Desplat$^2$}
\address{$^1$Department of Physics and Astronomy, 
University of Edinburgh, JCMB King's Buildings, Mayfield Road, Edinburgh\\
EH9 3JZ, United Kingdom\\
$^2$Edinburgh Parallel Computing Centre,
University of Edinburgh, JCMB King's Buildings, Mayfield Road, Edinburgh\\
EH9 3JZ, United Kingdom}
\maketitle

\vspace{-0.4cm}
\center{\small{(Draft printed \today)}}
\begin{abstract}
The persistence exponent, $\theta$, is defined by $N_F\sim t^{\theta}$, where
$t$ is the time since the start of the coarsening process and
the ``no-flip fraction'', $N_F$, is the number of points that have not
seen a change of ``color'' since $t=0$.
Here we investigate numerically the
persistence exponent for a binary fluid system where the coarsening
is dominated by hydrodynamic transport.  We find that $N_F$
follows a power law decay (as opposed to exponential) with the 
value of $\theta$ somewhat dependent on the domain growth rate
($L\sim t^{\alpha}$, where $L$ is the average domain size), in the
range $\theta=1.23\pm0.1$ ($\alpha = 2/3$) to $\theta=1.37\pm0.2$ ($\alpha=1$).
These $\alpha$ values correspond to the inertial and viscous hydrodynamic
regimes respectively.\hfill
{\rule[0ex]{4ex}{0ex}}
\end{abstract}
\bigskip
\footnotesize{\hspace{2cm}PACS numbers: 64.60.Ht, 64.75+g, 82.20.Wt}\hfill
\bigskip
]

\section{Introduction}
\label{sec:intro}

Persistence exponents can be defined for systems where there is an
order parameter whose time evolution can be followed at each point,
${\mathbf x}$, in the system.
It is most easily understood for an order parameter, $\phi({\mathbf x})$
that takes just two equilibrium values (spin up/down),
but the concept is easily generalised
({\textit e.g.} red/blue fluid for binary fluid mixtures).
The density of points for which sgn[$\phi({\mathbf x}) - \langle\phi\rangle$]
has not changed sign up to time $t$ as the system coarsens,
the ``no-flip fraction'', $N_F$, will, for a conserved order parameter,
or a quench to zero temperature, typically decay as a
power law, $N_F\sim t^{-\theta}$, where $\theta$ is the persistence exponent.

Persistence exponents were first investigated by Bray, Derrida and Godr\`{e}che
\cite{bray94b,derrida94a}, in the context of one dimensional diffusive systems.
For example, for the $q$-state Potts model in one dimension,
Derrida {\textit et. al.}
\cite{derrida94a} found from simulations, and later proved analytically
\cite{derrida95a}, that $\theta=3/8$ for
$q=2$, $\theta\simeq0.53$ for $q=3$ and $\theta\rightarrow 1$ as
$q\rightarrow\infty$.
It is also possible to define persistence
exponents for systems with non-conserved order parameters at finite
temperature by coarse-graining \cite{finitenc}.

As with most critical systems, analytical calculations are difficult;
a few exact results exist \cite{bray94b,derrida95a}.
Some mean field calculations have been done \cite{majumdar96a},
however, mean field approximations also predict that $\theta$ is not
independent of the other critical exponents.
Mean field theory assumes that the order parameter dynamics is a Markov
process, and Derrida {\textit et. al.} and Majumdar {\textit et. al.} \cite{derrida96a,majumdar96a} 
argue that in general this is not the case, and thus $\theta$ is, in fact,
a new critical exponent independent of the two static
and two dynamic exponents already known.

In this paper we present an investigation of the persistence behavior of
our simulation of 3D spinodal decomposition in a binary fluid mixture,
other results of which are reported elsewhere \cite{kendon99a,kendon99b}.
Brief details of the theoretical model and simulation method are given first,
followed in Section \ref{sec:theory} by a theoretical approach to
persistence behavior in this system.
In Section \ref{sec:analysis} the numerical analysis is described, with
the results for the flip rate given in Section \ref{sec:fliprate} and
estimates of the persistence exponent in Section \ref{sec:results}.
Finally, in Section \ref{sec:conc}, the results are summarised.

\section{Model and simulation}
\label{sec:model}

A symmetric binary fluid mixture differs from a purely diffusive system
such as an alloy in that the phase separation is assisted by hydrodynamics.
There is an initial diffusive period during which an interlocking structure
of single-fluid regions is formed, separated by sharp interfaces.
The interfaces then take over as the driving force, displacing the
fluid as they flatten and shrink, leading to a much more rapid domain growth
rate of $L\sim t$ (in 3D) compared to diffusive growth of $L\sim t^{1/3}$,
where $L$ is the average domain size.
This linear growth was first predicted by Siggia \cite{siggia79a}.
Furukawa \cite{furukawa85a} later predicted that as the inertial effects
came into play, the growth rate would slow to $L\sim t^{2/3}$.
These growth rates have been observed in simulation
\cite{kendon99a,footsim}
and linear growth has been observed experimentally \cite{experiments}.
Recent suggestions that the growth rate may slow still further
to $L\sim t^{1/2}$ (or slower) \cite{grant99a} are not supported by
our simulation work \cite{kendon99a} or theory \cite{kendon99c}, and
will not concern us here.

For numerical work, we use the following model free energy:
\begin{equation}
F = \int d{\mathbf r}\left\{-\frac{A}{2}\phi^2+\frac{B}{4}\phi^4
  + \tilde\rho\ln\tilde\rho +
\frac{\kappa}{2}|\nabla\phi|^2\right\},
\end{equation}
in which $A$, $B$ and $\kappa$ are parameters that determine
the interfacial width ($\xi=\sqrt{\kappa/2A}$),
and interfacial tension ($\sigma = \sqrt{8\kappa A^3/9B^2}$);
$\phi$ is the usual order parameter (the normalized difference in number
density of the two fluid species);  $\tilde\rho$ is the
total fluid density, which remains (virtually) constant
throughout \cite{swift96a,ladd94a}.
We choose $A/B = 1$ so $\phi = \pm 1$ in equilibrium.
Our simulation \cite{bladon99a} uses a lattice Boltzmann (LB) method
\cite{higuera89a,swift96a} with a cubic lattice with nearest and
next-nearest neighbor interactions (D3Q15).
It was run on Cray T3D and Hitachi SR-2201 
parallel machines with system sizes up to $256^3$.

\begin{figure}[tb!]
    \resizebox{0.46\textwidth}{!}{\rotatebox{-90}{\includegraphics{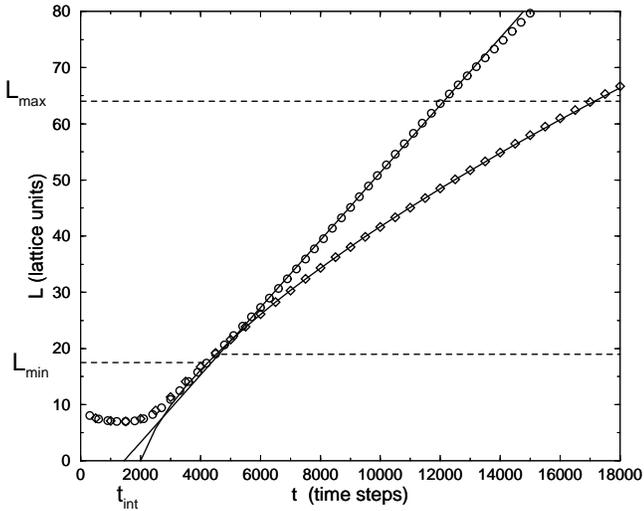}}}
    \caption{$L$ vs. $t$ for the runs with $L_0=5.9$ (circles)
             and $0.0003$ (diamonds).  The region used for fitting is
	     delimited by $L_{\mathrm min} < L < L_{\mathrm max} = 64$,
	     and the fits (solid) (to $\alpha=1$ and $\alpha=2/3$ respectively)
	     are projected back to show the intercepts, $t_{\mathrm int}$.}
    \protect\label{graph:raw-data}
\end{figure}

The data used for the analysis of persistence behavior corresponds to 
that used in our earlier studies of the growth exponents \cite{kendon99a}.
Details of the simulation parameters for each run, and selection
of usable data can also be found in \cite{kendon99a}.
The filters used to eliminate diffusive and finite size effects
mean that the good data from any single $256^3$
run lies within $20 \lesssim L \le 64$ in units of the lattice spacing.
The data was fitted to $L=b(t-t_{\mathrm int})^{\alpha}$, where
$b$ is a prefactor determined by the physical parameters and
$t_{\mathrm int}$ is a non-universal adjustment to the zero point on the 
time scale dependent on the initial diffusive period. 
Figure \ref{graph:raw-data} shows two sample runs and the fitted curves.
The values of $\alpha$ and $t_{\mathrm int}$ will be used in the subsequent
analysis of the persistence behavior.
By using characteristic length ($L_0=\eta^2/(\rho\sigma)$) and time
($t_0=\eta^3/(\rho\sigma^2)$) scales uniquely defined by the physical
parameters, the data from all the runs was combined into a single
$L$, $t$ plot covering a linear region, through a broad crossover, to $t^{2/3}$
\cite{kendon99a}.
The value of the growth exponent, $\alpha$, where the domain size,
$L\sim t^{\alpha}$, was found to range from 1.0 to 0.67, thus the
data spans the full range from viscous to inertial hydrodynamic growth.
The breadth of the crossover region justifies the use of a single effective
exponent, $\alpha$, to fit any single run.
It is convenient to use the value of $L_0$ (in simulation units) to
distinguish the different simulation runs.  The viscous regime corresponds
to $L_0 > 1$, the inertial regime to $L_0 < 0.001$, with intermediate
values corresponding to the crossover region.

\section{Flip-rate model}
\label{sec:theory}

For the 3D binary fluid system it is unrealistic to expect to derive
anything but the simplest approximate results using a theoretical
approach.  However, this will serve to illuminate the persistence-related
quantities under discussion.
The simple results to be derived in this section will then be of
assistance in the numerical analysis that follows.

Consider the step by step ($\Delta t=1$ in simulation units)
time evolution of the value of the order parameter at
a single point as the system phase separates and coarsens from the
initially completely mixed state.  A typical example is represented
schematically in Fig. \ref{fig:persist_line}.
In order to focus on the behavior in the hydrodynamic regime, 
an initial state is chosen at a reference point, $t_{\mathrm start}$, 
a time corresponding to when domains have coarsened to  a size
$L \sim L_{\mathrm min}$ that marks the onset of purely hydrodynamic
behavior (determined by the point at which diffusive
growth has fallen below 2\% \cite{kendon99a}).  The ``no-flip fraction'',
$N_F(t/t_{\mathrm start})$ is then defined as the fraction of sites that
have not changed color since $t_{\mathrm start}$.
Scaling by $t_{\mathrm start}$ is equivalent to choosing units such
that $t_{\mathrm start}=1$, which is in any case the value of the initial
reference time used in other studies.

In order to derive an approximate functional form for
$N_F(t/t_{\mathrm start})$, it is useful to define two further quantities,
\setlength{\parskip}{0.0ex plus 0.5ex minus 0.2ex}
\begin{itemize}
\item the flip rate, $P_F(t)$ is simply the
proportion of sites that changed color between time step $t-1$ and $t$, and,
\item the flip probability, $P(t, t_1)$, is the probability that a site
changes color at time $t$,
{\textit given that it last changed color at time $t_1$}.
\end{itemize}

These two quantities are related as follows:
\begin{equation}
P_F(t) = \int_{t_{\mathrm int}}^t\!{\mathrm d}t_1\,P_F(t_1)\,P(t,t_1),
\label{eq:pers1}
\end{equation}
where the sum over discrete time in our simulations has been 
approximated by an integral.
Equation (\ref{eq:pers1}) says that the flip rate at time $t$ is given by 
all the points that last flipped at time $t_1$ and are due to flip again 
at time $t$, {\textit i.e.} $P(t,t_1)$, integrated over all possible prior flip 
times, $t_{\mathrm int}<t_1<t$, and weighted by the number of sites with prior 
flip time $t_1$, {\textit i.e.} $P_F(t_1)$.
The lower limit of the integral is set to $t_{\mathrm int}$ because,
as can be seen from Fig. \ref{graph:raw-data}, $t_{\mathrm int}$
corresponds to the natural zero point on the time scale, the time at which,
(ignoring diffusion and the finite width of the interfaces),
the domain size would be zero so that, in effect, every site flipped at
$t=t_{\mathrm int}$.  Henceforth $t_{\mathrm int}$ will be usually be set
to zero, to simplify the algebra.

Equation (\ref{eq:pers1}) is an integral equation in two unknown functions,
$P(t,t_1)$ and $P_F(t)$.  A solution for $P(t,t_1)$ can be obtained by
making some assumptions about the asymptotic form of the simpler quantity,
$P_F(t)$, for the 3D hydrodynamic system.
The dynamics are determined by the basic scaling growth law, 
$L=b(t-t_{\mathrm int})^{\alpha}$, where the prefactor $v$ depends on
the system parameters (density, viscosity, interfacial tension).
Once the initial diffusive period is over, and the domains of
red and blue fluid are separated by well-formed interfaces,
lattice sites change color from red to blue when an interface moves across
them as the domains grow in size.
The flip rate, $P_F(t)$, is thus
given by the rate at which the interface moves through the system sweeping
out a volume of points that change color.  
So, an estimate of the area of the interface, and its 
average speed, will yield an estimate for $P_F(t)$.
This is the ``flip-rate model''.

The area of the interface is given approximately 
by $A(t) = c_L V/L$ where $V$ is the system volume
and $c_L$ is a prefactor of order unity.
As the system coarsens, the interfaces must move to accommodate
the enlargement of the domains.
It has been shown \cite{kendon99b} that the average fluid
velocity is comparable in magnitude with ${\mathrm d}L/{\mathrm d}t$,
so it is reasonable to estimate the speed of the interface as
$c_v\,{\mathrm d}L/{\mathrm d}t$, where $c_v$ is another prefactor of order unity.
The volume swept out per unit time will
be $A(t)\,c_v\,{\mathrm d}L/{\mathrm d}t$. 
Combining both $c_L$ and $c_v$ into a single prefactor, $c$,
the flip rate per unit volume will be approximately,
\begin{equation}
P_F(t) = c\frac{1}{L}\frac{{\mathrm d}L}{{\mathrm d}t} = \frac{c\alpha}{t}.
\label{eq:pers2}
\end{equation}
This approximation for $P_F(t)$ diverges at $t=0$.  In the discrete time
of our simulations, we start with timestep $t=1$ so no divergences arise;
in the continuum approximations it is appropriate to start at time $t=0$,
thus we will need to take care that no integrals diverge at their
lower limits.

Substituting $P_F(t) = c\alpha/t$ into Eq. (\ref{eq:pers1}) gives,
\begin{equation}
\int_0^t{\mathrm d}t_1\,\frac{t}{t_1}P(t,t_1) = 1,
\label{eq:pers3}
\end{equation}
where have set $t_{\mathrm int} = 0$.
Equation (\ref{eq:pers3}) has solutions for $P(t,t_1)$ of the form,
\begin{equation}
P(t,t_1) = (\beta-1)t_1^{\beta-1}t^{-\beta}, \mbox{\hspace{1cm}} \beta > 1,
\label{eq:pers4}
\end{equation}
where $\beta$ is an arbitrary exponent, as can readily be 
verified by substitution.
We have $P(t,t_1)\rightarrow 0$ for $t_1 \rightarrow 0$ so the integral is
well-behaved at $t_1=0$.
The condition,
\begin{equation}
\int_{t_1}^{\infty}{\mathrm d}t\,P(t,t_1) = 1,
\label{eq:pers5}
\end{equation}
({\textit i.e.} all sites do, eventually, flip)
is also satisfied, so $P(t,t_1)$ is a properly normalised probability.


An expression can now be written down for the ``no flip fraction'',
$N_F(t/t_{\mathrm start})$, which is the fraction of sites that
have not changed color since $t_{\mathrm start}$,
\begin{equation}
N_F = \int_0^{t_{\mathrm start}}{\mathrm d}t_1\,\int_t^{\infty}{\mathrm d}t_2\,
        P(t_2,t_1)P_F(t_1),
\label{eq:pers6}
\end{equation}
{\textit i.e.} we count every point whose last flip was before $t_{\mathrm start}$ and whose
next flip is after time $t$.  Substituting for $P_F$ and $P(t_2,t_1)$ gives,
\begin{eqnarray}
N_F&=&c\alpha\int_0^{t_{\mathrm start}}\!{\mathrm d}t_1\,(\beta
-1)t_1^{\beta-2}
        \int_t^{\infty}\!{\mathrm d}t_2\,t_2^{-\beta}, \nonumber \\
   &=&\frac{c\alpha}{\theta}\left(\frac{t_{\mathrm start}}{t}\right)^{\theta},
        \mbox{\hspace{1cm}} \theta>0,
\label{eq:pers7}
\end{eqnarray}
setting $\theta = (\beta-1)$ as the persistence exponent.
The flip-rate model has thus provided an expression for 
$N_F$ that has the exponent, $\theta$, in the prefactor 
as well as being the asymptotic power of the decay.
This puts helpful constraints on the data analysis,
although for it to be really useful, the prefactor, $c$,
in Eq. (\ref{eq:pers2})
has to be pinned down as precisely as possible.
In a log-log plot,
\begin{equation}
\log N_F = \log (c\alpha/\theta) - \theta\log(t/t_{\mathrm start}).
\label{eq:pers8}
\end{equation}
so plotting $N_f$ against $t/t_{\mathrm start}$ should give a straight line
when viewed log-log, with slope $-\theta$, for $t\gg t_{\mathrm start}$.
The initial data near $t\gtrsim t_{\mathrm start}$ will depart from
this asymptotic behavior and tend to $N_F=1$ because by
definition nothing flipped yet for $t = t_{\mathrm start}$.  The fit to the
asymptotic slope should, however, have an intercept at $c\alpha/\theta$
for $t=t_{\mathrm start}$.


It is possible to go on to define a hierarchy of persistence
quantities, of which $N_F$ is the first \cite{derrida94a}.
The ``one flip fraction'', $O_F$ is the density of points that have
changed color exactly once since time $t_{\mathrm start}$.
Within the flip-rate model, an expression for $O_F$ for this
system can be derived in the same way as for $N_F$.  
\begin{eqnarray}
&&O_F(t/t_{\mathrm start}) = \nonumber\\
&&                \int_t^{\infty}\!\!\!{\mathrm d}t_2\,
                  \int_{t_{\mathrm start}}^{t_2}\!\!\!{\mathrm d}t_f\,
                  \int_0^{t_{\mathrm start}}\!\!\!{\mathrm d}t_1\,
                            P(t_2,t_f)P(t_f,t_1)P_F(t_1).\nonumber\\
\label{eq:pers9}
\end{eqnarray}
The ``one flip'' occurs at time $t_f$ with $t_{\mathrm start}<t_f<t_2$.
The probability of an initial flip at $t_1$
with $0<t_1<t_{\mathrm start}$, then flipping at time $t_f$ is
$P(t_f,t_1)$.  The probability of flipping again at time $t_2$ is then
$P(t_2,t_f)P(t_f,t_1)$, provided the two flip events are independent
of each other (so the probabilities can be multiplied).
In other words, this model assumes that the flip probability, $P(t_2,t_f)$
is only dependent on the last time the site flipped,
it has no memory of earlier flips.  This requires a Markov process for the
coarsening dynamics, which Derrida {\textit et. al.} and
Majumdar {\textit et. al.} \cite{derrida96a,majumdar96a}
claim is not generally the case for the non-hydrodynamic systems they
have considered.
This therefore constitutes another untested assumption in the flip-rate
model described here.

Substituting for $P_F$ and $P(t_2,t_1)$ as before gives,
\begin{eqnarray}
O_F&=&\int_t^{\infty}\!\!\!{\mathrm d}t_2\,\int_{t_{\mathrm start}}^{t_2}\!\!\!{\mathrm d}t_f
\,
        \int_0^{t_{\mathrm start}}\!\!\!{\mathrm d}t_1\,\frac{c\alpha}{t_1}\,
        \theta\frac{t_f^{\theta}}{t_2^{\theta+1}}\,
        \theta\frac{t_1^{\theta}}{t_f^{\theta+1}} \nonumber \\
   &=&\frac{c\alpha}{\theta}\left(\frac{t_{\mathrm start}}{t}\right)^{\theta}
        \left[1 - \ln\left(\frac{t_{\mathrm start}}{t}\right)^{\theta}\right],
\label{eq:pers10}
\end{eqnarray}
after integrating by parts.  This has a dominant logarithmic term
so doesn't have an asymptotic power law decay.
(In contrast, in the 1D Ising model, $O_F$ appears to have
the same asymptotic behavior as  $N_F$, see \cite{derrida94a,derrida95a}.)
This could, of course, be a result of the limitations of the flip-rate model,
however, the simulation results suggest that the asymptotic
behavior of $O_F$ is different from that of $N_F$ (see Figs. 
\ref{graph:N_F} and \ref{graph:O_F} below).
Again, this expression is only valid for $t\gg t_{\mathrm start}$ since 
$O_F\rightarrow 0$ for $t\rightarrow t_{\mathrm start}$.  As with $N_F$, the
(same) critical exponent $\theta$ appears both as an exponent and
as a prefactor so providing an extra constraint on any fits to
simulation data.

Nothing in the theory so far is specific to three space dimensions.
However, in two dimensional binary fluid systems, the interface
doesn't completely interconnect \cite{wagner98a},
so the approximation $P_F=c\alpha/t$ is unlikely to work without 
modification.  The two dimensional form of the theory will not
be pursued further here.

\section{Numerical analysis}
\label{sec:analysis}

The flip-rate model just described will now be used to guide the
analysis of the data from our spinodal decomposition simulations.
The data from the largest ($256^3$) runs is typically saved every 200
to 500 time steps, $\Delta t_{\mathrm stride}$.
It is also coarse-grained from a $256^3$ lattice
down to $128^3$ by averaging over groups of eight neighboring lattice points.  
This data can be used to calculate the four key quantities discussed in
the previous section, $P_F$, $P(t_1,t_2)$, $N_F$ and $O_F$, provided 
the spatial and temporal coarse-graining makes no significant
difference to the results.

\begin{figure}[tb!]
    \resizebox{0.46\textwidth}{!}{\rotatebox{-90}{\includegraphics{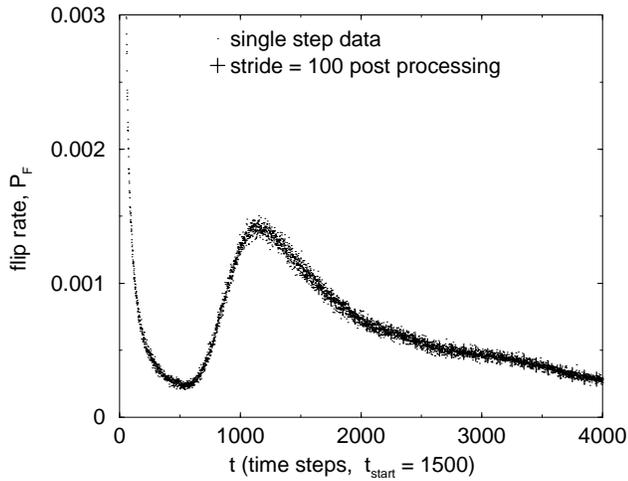}}}
    \caption[Flip rate, $P_F$, under temporal coarse-graining]
            {Flip rate, $P_F$, for $96^3$ system single step
             (dots) and stride=100 (crosses).}
    \protect\label{graph:P_F_96}
\end{figure}

To investigate temporal coarse-graining, a special run on a $96^3$ grid was done
with the key quantities calculated every time step.  This extra computation
significantly slows down the computer run time so it isn't practical
to perform larger simulations with single-step calculations.
The results from a single-step
calculation were then compared with the same run analysed over a stride of
100 time steps, see Fig. \ref{graph:P_F_96}.
The result of this comparison showed excellent agreement between
the single-step and strided estimates of $P_F(t)$.

In fact, the errors introduced by analysing only every $\Delta t_{\mathrm stride}$
steps rather than every single step arise from miss-counting sites that
flip more than once in that period.
If the flip rate, $P_F(t)$, is small then the number of
multiple flips will be much smaller still so the errors will negligible.
In all runs analysed, $P_F$, was found to be small 
even over the largest $\Delta t_{\mathrm stride}$ used,
{\textit i.e.} $P_F\times \Delta t_{\mathrm stride} \ll 1$ (a value of
one would mean every site flipped).

The shape of $P_F(t)$ in Fig. \ref{graph:P_F_96}
for times $t<t_{\mathrm start}$ clearly shows a transition
from diffusive to hydrodynamic behavior in the initial fall (diffusion),
rise (interfaces start to move) and fall again (hydrodynamic).
Thus $P_F(t)$ is a sensitive indicator of the system dynamics, and
the choice of $t_{\mathrm start}=1500$ in this $96^3$ system is confirmed to be
located, as desired, near the beginning of the hydrodynamic regime but not
too close to the transition from diffusive behavior.

The effect of spatial coarse-graining was checked by comparing the
results of analysing strided data from $128^3$ and $256^3$ runs
with identical parameters apart from system size.
Excellent agreement was found everywhere except
in the tail of $P(t,t_{\mathrm start})$, which can be explained
by poor statistics in this region.
The conclusion, 
therefore, is that spatial and temporal coarse-graining on the scale
used in this analysis does not introduce significant errors into the results.

From the main simulation data, good statistics for $P_F$, $N_F$ and $O_F$
can be obtained, as every lattice point can be assessed according to if/when
it next flips after $t_{\mathrm start}$.  The statistics are not so good
for estimating $P(t,t_{\mathrm start})$, because the reference base is
reduced to only those sites that flip between $t_{\mathrm start}$ and
$\Delta t_{\mathrm stride}$ rather than the whole system;
recall that $P(t,t_{\mathrm start})$ refers to the first
flip happening exactly at $t_{\mathrm start}$.

\section{Flip rate}
\label{sec:fliprate}

\begin{figure}[tb!]
    \resizebox{0.46\textwidth}{!}{\rotatebox{-90}{\includegraphics{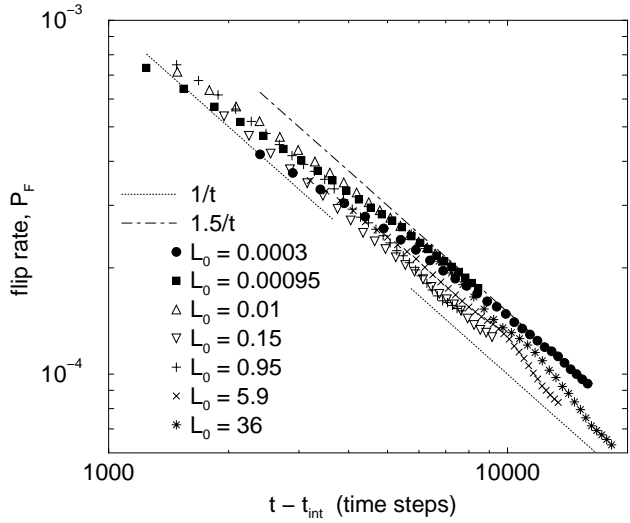}}}
    \caption{Flip rate, $P_F$, for simulation runs as identified by
	     $L_0$ values in key, log-log plot.
	     Also shown for comparison, $1/t$ and $1.5/t$.}
    \protect\label{graph:flip_rate_ll}
\end{figure}

In order to make full use of the model presented in
Section \ref{sec:theory}, it is
necessary to investigate whether the flip rate really
follows the approximate theoretical expression, $P_F = c\alpha/t$,
Eq. (\ref{eq:pers3}), and whether the prefactor, $c$,
can be evaluated sufficiently accurately from the simulation data
to allow a strongly constrained fit to be done for $\theta$ in the
analysis of $N_F$ and $O_F$.
Figure \ref{graph:flip_rate_ll} shows $P_F$ for each simulation run
in a log-log plot, with $1/t$ and $1.5/t$ also shown for comparison.
The prefactor, $c$, clearly varies somewhat over time, and with $\alpha$,
although the variation with time can partly be explained since early times,
$t \sim t_{\mathrm start}$, are expected to differ from asymptotic behavior.  

\begin{figure}[tb!]
    \resizebox{0.46\textwidth}{!}{\rotatebox{-90}{\includegraphics{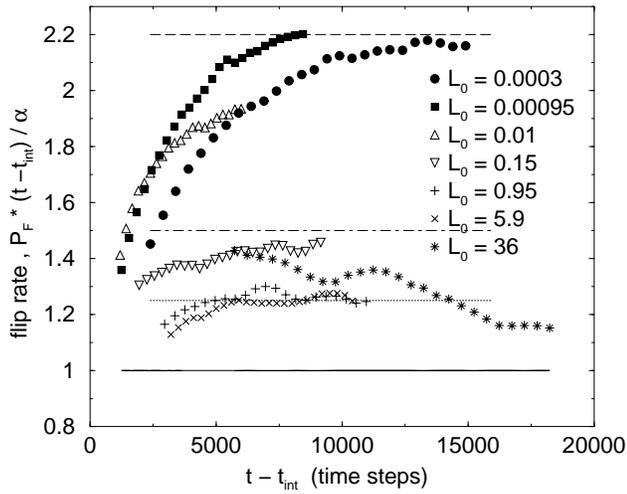}}}
    \caption{Flip rate, $P_F$, rescaled by $(t - t_{\mathrm int})/\alpha$ on a
             linear plot so values of the prefactor, $c$, can be read
             off the ordinate axis.}
    \protect\label{graph:flip_rate}
\end{figure}

To determine the values and variation in $c$ more accurately,
a linear plot of the flip rate, $P_F$, scaled by
$(t-t_{\mathrm int})/\alpha$ is shown in Fig. \ref{graph:flip_rate}.
It can be seen that the results roughly split into two groups corresponding
to the runs found to be in the viscous and lower crossover regions
($L_0 \gtrsim 0.1$), with prefactors around $c\simeq 1.25$,
and runs in the inertial regime ($L_0 \lesssim 0.01$), with $c\simeq 2.2$.

If this difference in the value of $c$ is real, it implies that the
geometry of the separating domains is different between the
viscous and inertial regimes.  Any such difference is not apparent
to the eye in visualisations of the interface \cite{kendon99e}.
There is some evidence from analysis of the structure factor \cite{kendon99b}
that there is a structural difference in the domain
structure, but it is unclear whether this is enough to account 
for the observed difference in the value of $c$.
However, there is another possible interpretation of the data, which is
that the flip rate actually exceeds $c\alpha/t$ for the inertial regime runs.
For, once the viscosity is low enough for inertial effects to become 
significant, the interface may exhibit capillary excitations.
This oscillation of the interface could increase the measured flip rate
as sites near the interface repeatedly flip back and forth, but without
contributing to the sweeping that removes sites from the ``no flip'' category.  
If capillary excitations have a significant effect on the flip rate,
this part of the flip rate needs to be discounted in the
subsequent fitting to determine the persistence exponent.
Since the extent of any excess flip rate cannot be determined at this stage,
a variable prefactor, $c$, has therefore been carried through the fitting
process for $\theta$.

\section{Persistence exponent}
\label{sec:results}

The ``no flip fraction'', $N_F$, is the primary quantity of interest
in this work, being the quantity from which the persistence exponent
$\theta$, is defined, $N_F \sim t^{-\theta}$.
The simulation results for $N_F$, with the time scaled as
$(t-t_{\mathrm int})/t_{\mathrm start}$
and points for times at which $L < L_{\mathrm min}$ or $L > 64$ removed,
are plotted in Fig. \ref{graph:N_F}.
The runs in the inertial regime have tails that fall to around $N_F=0.08$,
while the viscous runs stop (reach domain size $L=64$) at around $N_F=0.15$.
There is thus barely one decade within which to estimate the slope of the tail.
Although a simple linear fit cannot produce a very accurate result,
superficial inspection suggests that $\theta$ is slightly greater
than unity, and fairly reproducible.

If we could use a fixed value of the prefactor, $c$, with confidence, the fit
could be constrained to intercept at $c\alpha/\theta$, Eq. (\ref{eq:pers8}).
Instead, a family of fits was done \cite{kendon99e},
covering a range of values of the prefactor,
$c$, and these were compared with a similar family of fits for the
one flip fraction, $O_F$.  
As can be seen in Fig. \ref{graph:O_F}, $O_F$ at best falls to only $0.2$ by
the end of the simulation data, so cannot be regarded as having
reached asymptotic behaviour yet.  The $O_F$ data was therefore used only
to confirm the direction in which $N_F$ is tending towards its asymptote.
From this fitting procedure, a range of consistent values of
both $c$ and $\theta$ were obtained, these are summarised in
Table \ref{table:persist}.
\begin{table}
  \begin{center}
  \caption{Summary of results for $c$ and $\theta$.}
  \protect\label{table:persist}
  \begin{tabular}{@{}r@{}lccc@{}}
    \multicolumn{2}{c}{$L_0$} & $\alpha$ & $c$ & $\theta$  \\
    \hline
    0&.0003 & 0.67 & 1.5  -- 1.9 & 1.20 -- 1.23 \\
    0&.00095 & 0.67 & 1.5  -- 2.2 & 1.20 -- 1.32 \\
    0&.01 & 0.75 & 1.3  -- 1.9 & 1.25 -- 1.40 \\
    0&.15 & 0.80 & 1.0  -- 1.5 & 1.15 -- 1.24 \\
    0&.95 & 0.95 & 0.85 -- 1.4 & 1.20 -- 1.40 \\
    5&.9 & 1.0  & 0.85 -- 1.4 & 1.25 -- 1.50 \\
    36& & 1.0  & 0.85 -- 1.3 & 1.30 -- 1.55 \\
  \end{tabular}
  \end{center}
\end{table}

\begin{figure}[tb!]
        \resizebox{0.46\textwidth}{!}{\rotatebox{-90}{\includegraphics{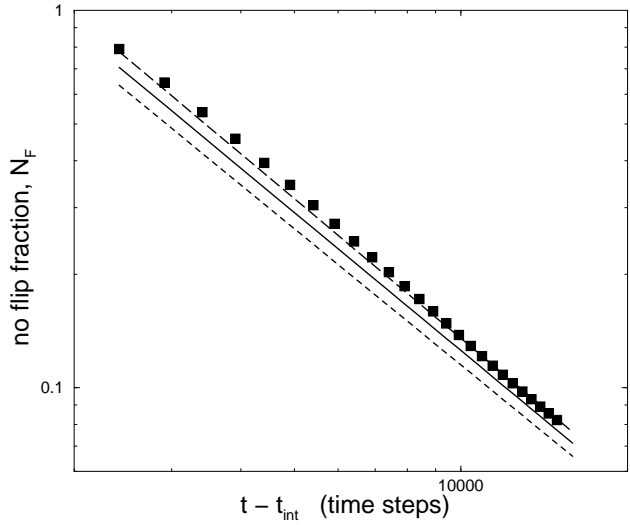}}} 
    \caption{Fitted curves for $N_F$ for the run with $L_0=0.0003$.
             The no flip fraction, $N_F$, (squares) calculated from
	     the numerical data is shown with the upper, median and
	     lower fits to Eq. (\protect\ref{eq:pers7})
             using $c=1.9$, $\theta=1.23$ (dashed)
	     $c=1.7$, $\theta=1.215$ (solid) and
	     $c=1.5$, $\theta=1.21$ (dotted) respectively
	     from Table \protect\ref{table:persist}.}
    \protect\label{graph:Run032_N_F}
\end{figure}

A sample plot for the run with $L_0=0.0003$ with $N_F$ and the
fitted expression in Eq. (\ref{eq:pers7}), with
high, median and low values of $c$ and $\theta$, is shown in 
Fig. \ref{graph:Run032_N_F}.

Fits to $P(t,t_{\mathrm start})=\theta/t(t_{\mathrm start}/t)^{\theta}$
were also done, producing values of $\theta$ consistent with those determined
from $N_F$ and $O_F$.  Since  $P(t,t_{\mathrm start})$ does not depend on
$c$ or $\alpha$, this is a useful check for consistency,
but because of the poorer statistics for $P(t,t_{\mathrm start})$,
no improvement in the range of values for $\theta$
was obtained by this extra step.

\begin{figure}[tb!]
        \resizebox{0.46\textwidth}{!}{\rotatebox{-90}{\includegraphics{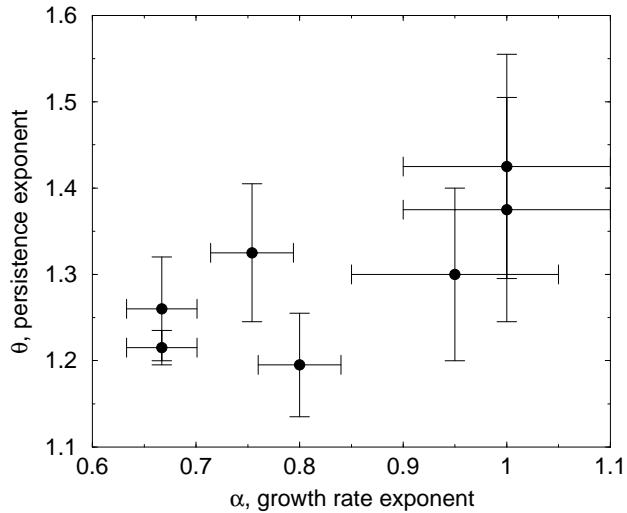}}}
    \caption{Persistence exponent, $\theta$, against spinodal
             growth rate exponent, $\alpha$, from simulation runs.
             The error bars are
             the range of $\theta$ from Table \protect\ref{table:persist},
             and the error estimates for $\alpha$
	     of 10\% (viscous regime) and 5\%
             (crossover and inertial regime)
             \protect\cite{kendon99b}.}
    \protect\label{graph:theta_alpha}
\end{figure}

The complete set of estimates for $\theta$ from Table \ref{table:persist}
is shown against the growth exponent, $\alpha$ in Fig. \ref{graph:theta_alpha},
with error bars indicating the range of values obtained
The results are consistent with
the persistence exponent, $\theta$, lying somewhere around 1.37 for
the viscous regime and 1.23 for the inertial regime.  
A single value of $\theta=1.333$ is also just about consistent with
the data, although some variation with $\alpha$ seems more likely.

The prefactor, $c$, could be as low as 1.0 for
viscous runs and rises to nearer 2.0 for inertial regime.
However, a value of $c=2.2$ as implied by the measured flip rate for
the inertial regime appears to be inconsistent with the 
persistence data for $N_F$, $O_F$, and $P(t,t_{\mathrm start})$.
The flip rate is thus sufficiently sensitive
to distinguish between viscous and inertial regime dynamics,
and suggests the presence of some mechanism such as capillary waves in
the inertial regime that raises the measured flip rate above that
predicted by the growth of the domains.

\section{Conclusions}
\label{sec:conc}

In this investigation of the persistence exponents for a 3D hydrodynamic
spinodal system, it has been observed that the decay of the
``no-flip fraction'', $N_F$, is a power law (as opposed to exponential).
It has not been possible to produce a very
accurate determination of the persistence exponent, $\theta$, 
but limits have been placed on the likely value, and it has been shown
that some variation with the effective growth exponent, $\alpha$, 
as one crosses slowly from viscous ($\alpha=1$) to inertial ($\alpha=2/3$)
coarsening, should be allowed for in any future studies.
The best estimate of $\theta$ for the viscous regime ($\alpha=1$), is
$\theta=1.37 \pm 0.2$.
From the point of view of fundamental
critical exponents, the value of $\theta$ in the inertial regime
is perhaps more significant, since this is (at least on current evidence)
the asymptotic behavior for spinodal decomposition, and here the
best estimate is $\theta=1.23 \pm 0.1$.
These represent the first estimates of $\theta$ in a system dominated by 
hydrodynamics rather than diffusion.
However, the error limits quoted here do not allow for systematic errors
arising from the use of the flip rate model itself.
More accurate determination 
from simulations would likely require significantly larger system sizes, 
which is difficult to envisage at present computing power.

\acknowledgements

We thank Alan Bray for drawing our attention to this problem, and
Martin Evans for valuable discussions.
(Work funded in part by EPSRC GR/M56234.)



%
\begin{onecolumn}
\begin{figure}[tbp!]
        \resizebox{\textwidth}{!}{\includegraphics{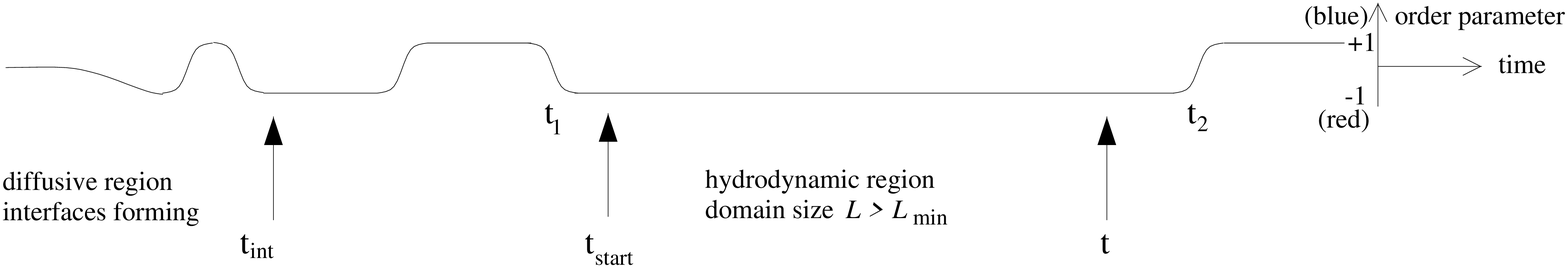}}
        \caption{Persistence timeline.}
        \protect\label{fig:persist_line}
\end{figure}
\end{onecolumn}
\begin{figure}[tb!]
        \resizebox{0.46\textwidth}{!}{\rotatebox{-90}{\includegraphics{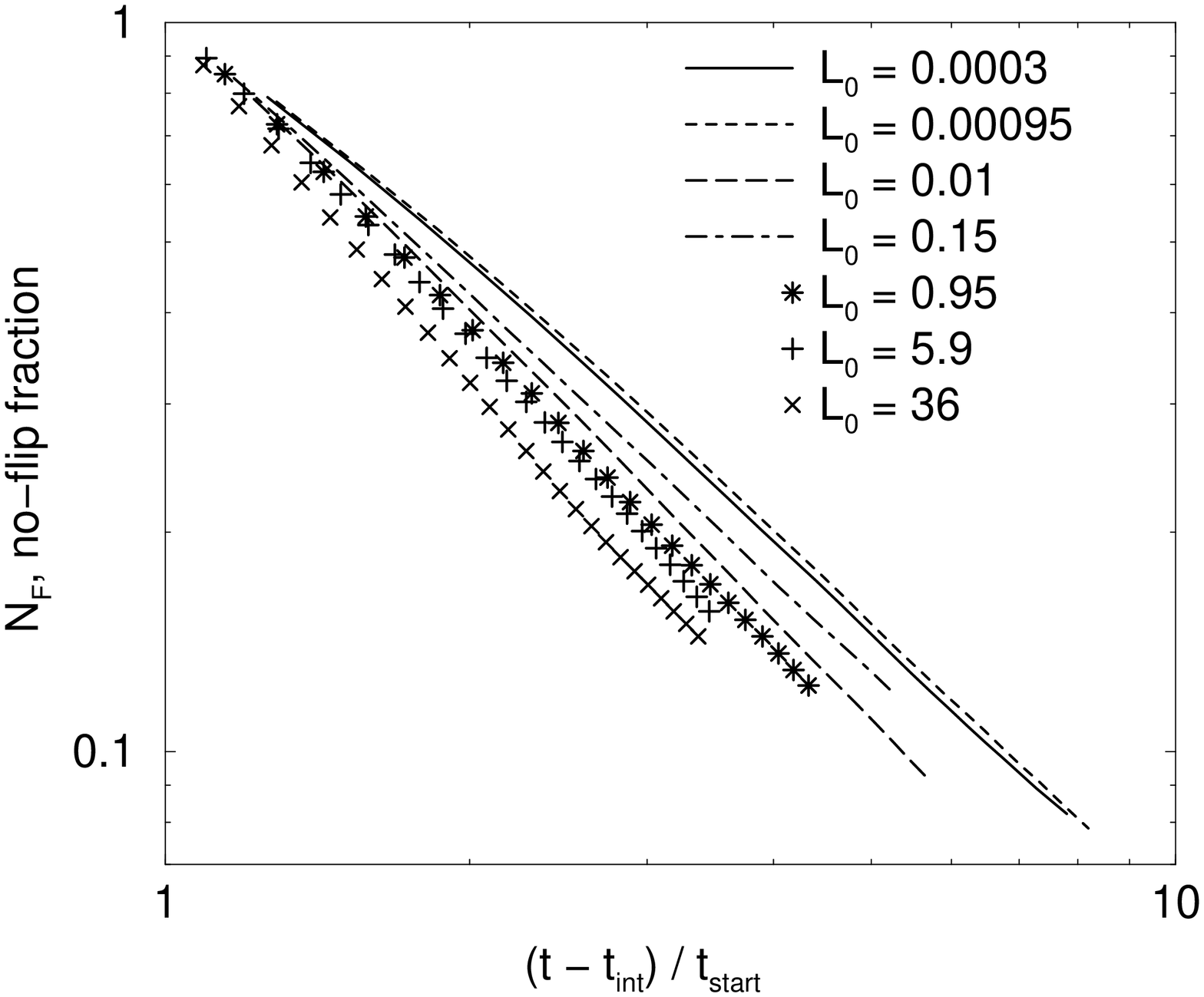}}}
    \caption{No-flip fraction, $N_F$, for all runs.}
    \protect\label{graph:N_F}
\end{figure}
%
\setlength{\textheight}{1.05\textheight}
\begin{figure}[tb!]
        \resizebox{0.46\textwidth}{!}{\rotatebox{-90}{\includegraphics{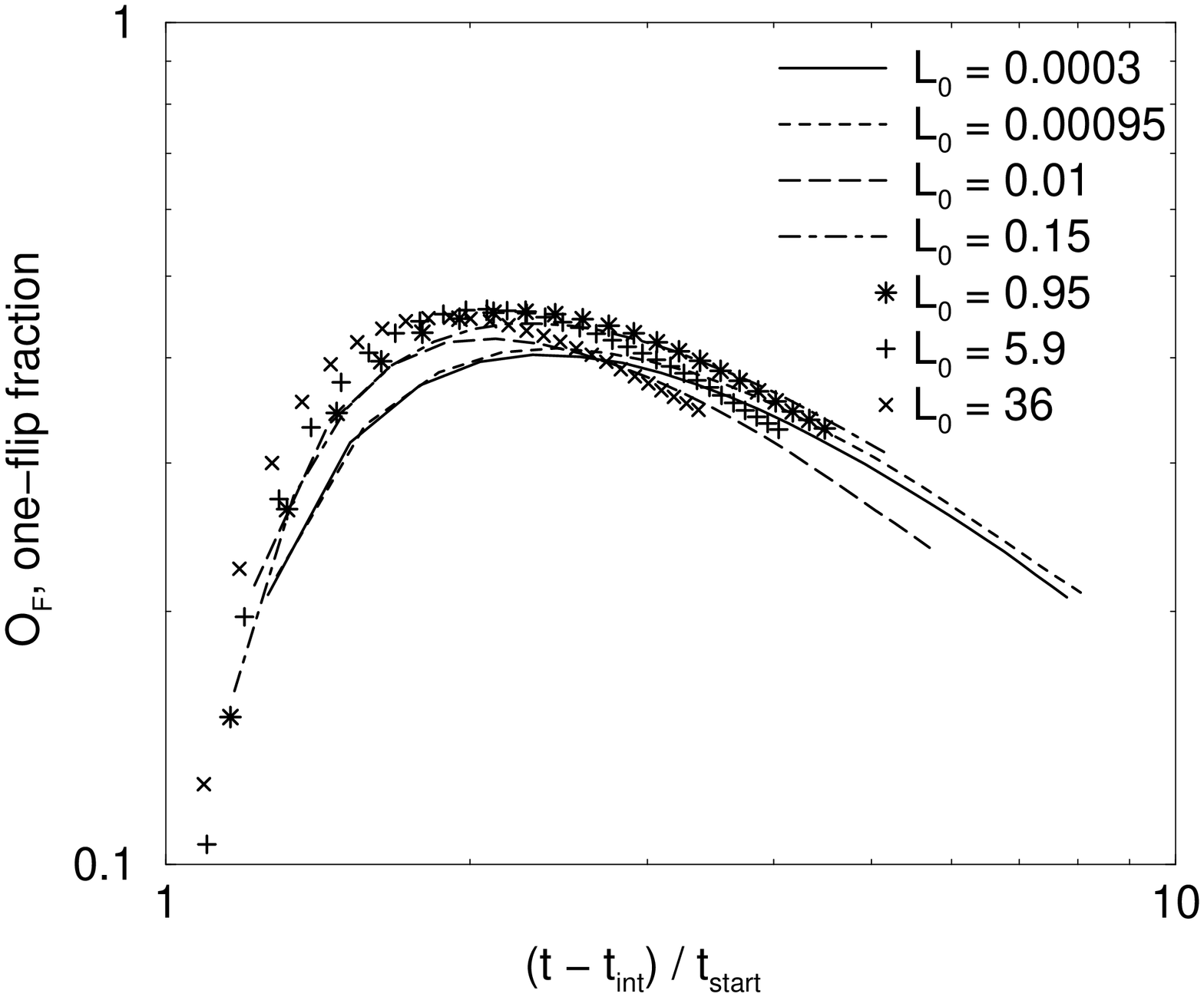}}}
    \caption{No-flip fraction, $O_F$, for all runs.}
    \protect\label{graph:O_F}
\end{figure}
%

\end{document}